\documentclass[aps,nofootinbib,superscriptaddress, showpacs,preprintnumbers,  nofootinbibt,twocolumn]{revtex4-2}
\usepackage{epsfig}
\usepackage{multirow}
\usepackage{subcaption}
\usepackage{eurosym}
\usepackage{dcolumn}
\usepackage{bm}
\usepackage{enumerate}
\usepackage{float}
\usepackage{epstopdf}
\usepackage{amsmath}
\usepackage{bm}
\usepackage{amsfonts}
\usepackage{amssymb}
\usepackage{graphicx}
\usepackage{alphalph,mathtools}
\usepackage{etoolbox}
\usepackage{color}
\usepackage[dvipsnames]{xcolor}
\usepackage{tensor}
\usepackage{amsthm}
\usepackage{booktabs}
\usepackage{hyperref}
\hypersetup{colorlinks,citecolor=blue}
\usepackage{footnote}
\usepackage{makecell,tabularx}

\def\be{\begin{equation}}
\def\ee{\end{equation}}
\def\bea{\begin{eqnarray}}
\def\eea{\end{eqnarray}}

\newcommand{\beq} {\begin{equation}}
\newcommand{\eeq} {\end{equation}}

\usepackage{relsize,exscale}

\begin{document}

\title{\bf Addressing the \(r_{d}\) Tension using Late-Time Observational Measurements in a Novel deceleration Parametrization}

\author{Himanshu Chaudhary}
\email{himanshuch1729@gmail.com}
\affiliation{Pacif Institute of Cosmology and Selfology (PICS), Sagara, Sambalpur 768224, Odisha, India.}
\affiliation{Department of Applied Mathematics, Delhi Technological University, Delhi-110042, India} 
\affiliation{Department of Mathematics, Shyamlal College, University of Delhi, Delhi-110032, India.}
\author{Ujjal Debnath}
\email{ujjaldebnath@gmail.com} 
\affiliation{Department of
Mathematics, Indian Institute of Engineering Science and
Technology, Shibpur, Howrah-711 103, India.}
\author{S. K. Maurya}
\email{sunil@unizwa.edu.om} 
\affiliation{Department of Mathematics and Physical Sciences, College of Arts and Sciences, University of Nizwa, Sultanate of Oman.}
\author{G.Mustafa}
\email{gmustafa3828@gmail.com } 
\affiliation{Department of Physics,
Zhejiang Normal University, Jinhua 321004, Peoples Republic of China}
\author{Farruh Atamurotov}
\email{atamurotov@yahoo.com}
\affiliation{Institute of Fundamental and Applied Research, National Research University TIIAME, Kori Niyoziy 39, Tashkent 100000, Uzbekistan}
\affiliation{University of Tashkent for Applied Sciences, Str. Gavhar 1, Tashkent 100149, Uzbekistan}
\affiliation{Shahrisabz State Pedagogical Institute, Shahrisabz Str. 10, Shahrisabz 181301, Uzbekistan}

\begin{abstract}
This paper introduces a novel cosmological model aimed at probing the accelerated expansion of the late Universe through a unique parametrization of the deceleration parameter. We aim to constrain key cosmic parameters by integrating recent measurements of the Hubble parameter obtained from various observational methods, including cosmic chronometers, Type Ia Supernovae, Gamma-Ray Bursts (GRB), Quasars, and Baryon Acoustic Oscillations (BAO) from recent galaxy surveys. With a redshift range spanning \(0.106 < z < 2.33\) and incorporating the latest Hubble constant measurement from Riess in 2022, our analysis yields optimal fit values for the Hubble parameter (\(H_{0}\)) and sound horizon (\(r_{d}\)). Notably, we uncover an inconsistency in \(H_{0}\) values derived from late-time observational measurements, reflecting the well-known \(H_{0}\) tension. In terms of \(r_{d}\), while there is close agreement between Joint analysis and Joint analysis with R22, discrepancies arise upon gradual inclusion of BAO and BAO with R22 datasets. Our model demonstrates excellent fit to observed data and aligns well with the standard \(\Lambda\)CDM paradigm at higher redshifts. However, its most intriguing aspect lies in predicting a super-accelerated expansion in the distant future, in contrast to the de Sitter phase predicted by \(\Lambda\)CDM. Additionally, unique behaviors in the jerk parameter hint at novel dynamics beyond traditional cosmological models. Statefinder and \(O_{m}\) Diagnostics tests were conducted, and comparison using the Akaike information criterion indicates neither model can be ruled out based on the latest observational measurements. These findings propose our cosmological model as a compelling alternative to \(\Lambda\)CDM, offering fresh insights into dark energy's nature and the cosmos' future.
\end{abstract}
\maketitle

\section{Introduction}
The discovery of the acceleration of cosmic expansion stands as one of the most significant breakthroughs in modern cosmology \cite{riess1998observational}. Instead of slowing down due to gravity's pull, the universe is actually expanding at an accelerating rate, driven by a mysterious force known as dark energy (DE) \cite{RevModPhys.75.559}. This enigmatic form of energy is believed to permeate all of space, pushing galaxies apart and causing the universe to stretch at an ever-increasing pace. In our current understanding, the composition of the Universe according to the classic cosmological cold dark matter (CDM) model is as follows: approximately 4.9\% is ordinary matter, 26.8\% is dark matter (DM), and a staggering 68.3\% is DE. DM, another cosmic enigma, reveals itself only through gravitational effects, as it neither emits nor absorbs light \cite{bertone2018history}. Despite its invisibility, its presence is evident from astronomical observations. The nature of both DE and DM remains elusive \cite{spergel2000observational,wang2016dark}, prompting ongoing investigations. While DE propels the expansion of the Universe, DM influences the formation and structure of galaxies and galaxy clusters. Together, they shape the cosmic landscape in ways that challenge our understanding of the fundamental forces governing the Universe.\\\\
Numerous theoretical frameworks have emerged in attempts to explain the nature of DE , each proposing different mechanisms to explain the observed accelerated expansion of the Universe. These approaches span from quintessence to holographic models, each with its own unique features and implications. The Quintessence suggests that DE is a dynamic field evolving over time, akin to a scalar field. This model introduces new physics beyond the cosmological constant hypothesis, allowing for variations in the DE density and equation of state (EoS) \cite{steinhardt2003quintessential}. The quintom model extends quintessence by considering scenarios where DE undergoes a transition between quintessence-like behavior and phantom-like behavior, where the EoS parameter crosses the boundary of -1, also known as the phantom divide \cite{wei2005hessence}. The Chaplygin Gas model, originally conceived as a means to reconcile the properties of DE and DM, has undergone modifications to include the effects of dark energy. This revised version presents a unified framework by introducing a generalized EoS capable of encompassing the behaviors of both DE and DM \cite{dev2003cosmological}. The K-essence models involve scalar fields with non-canonical kinetic terms. These models provide a flexible framework to generate a wide range of DE behaviors, including tracking solutions and scaling solutions \cite{bose2009k}. The New Age Graphic model introduces a DE parameterization based on the concept of graphical DE, allowing for deviations from the standard cosmological constant model while remaining consistent with observational data \cite{wei2008new}. The Holographic DE Model inspired by the holographic principle, this model relates the DE density to the future event horizon of the universe. It suggests that DE is a manifestation of the underlying quantum structure of spacetime \cite{wang2017holographic}. The Pilgrim DE model incorporates an interaction between DE and DM , offering a mechanism for alleviating the coincidence problem through a time-varying DE EoS \cite{sharif2013analysis}. The Tsallis Holographic DE combines the holographic principle with the Tsallis entropy to formulate a new approach to DE. It explores the connection between DE and non-extensive statistical mechanics, offering insights into the thermodynamic properties of the Universe \cite{saridakis2018holographic}. Despite the diversity of these models, the simplest explanation for DE remains the cosmological constant hypothesis, consistent with observational data \cite{visinelli2019revisiting}. However, this hypothesis faces several challenges, including the coincidence problem (the question of why the density of DE is of the same order as the density of matter in the universe), the fine-tuning problem (the problem of explaining why the cosmological constant is so small), and the age problem (the tension between the age of the universe inferred from observations and the age predicted by the cosmological constant model) \cite{arkani2000new,quartin2008dark,chen2012age}. These unresolved issues continue to drive research into alternative theories of DE, pushing the boundaries of our understanding of the cosmos.\\\\
The EoS parameter $\omega$ defines the relationship between energy density and pressure in the cosmic framework. This is a dimensionless parameter that describes the cosmos' phases \cite{caldwell1998cosmological}. A parametrized version of $\omega$ is assumed for the departing behavior of DE. A Taylor series expansion in the redshift, scale factor or any other parametrization of may be used \cite{visser2004jerk}. The main idea in this approach is to consider a specific evolution scenario instead of considering any DE model apriori and then determine the nature of the exotic component that is triggering cosmic acceleration. It is known as the model-independent approach, which depends on estimating model parameters from existing observational datasets \cite{del2012three,katore2015anisotropic,riess2004type,bouali2023data,bouali2023constraints,bouali2023cosmological}. However, number of studies have tried to find out  the history of the Universe without relying on any particular cosmological model. Such techniques are frequently referred to as cosmography or cosmokinetic models, but we shall simply refer to them as kinematic models \cite{capozziello2019kinematic,
benetti2019connecting}. This name stems from the fact that the entire study of the Universe's expansion (or its kinematics) is characterised solely by the Hubble expansion rate $H = \frac{\dot{a}}{a}$, the deceleration parameter $q = -\frac{a\ddot{a}}{\dot{a}^2}$ and the jerk parameter $j = -\frac{\stackrel{\ldots}{a}a^{2}}{\dot{a}^3}$, where a is the scale factor in the (Friedmann-Lema\^itre-Robertson-Walker (FLRW) metric. The deceleration parameter allows us to investigate the transition from a decelerated to an accelerated phase, whereas the jerk parameter allows us to investigate deviations from the cosmic concordance model without being limited to a certain model. In terms of the deceleration parameter, various studies have attempted to determine the redshift at which the Universe transitions to an accelerated phase. \cite{mamon2018constraints,
akarsu2014probing,
xu2008constraints}. In modern cosmology, A model-independent determination of the current deceleration parameter $q_0$ and the deceleration-acceleration transition redshift are two of the most important things. It is required to utilize some parametrization for the deceleration parameter in order to investigate it in a cosmological-model-independent context. This system has both benefits and drawbacks. One benefit is that it does not depend on the amount of matter and energy in the Universe. One problem with this method is that it is unable to clarify the reason for accelerating expansion. Furthermore, the current deceleration parameter's value may be affected by the anticipated form of $q(z)$. Recently, \cite{mamon2017parametric} studied a special form of deceleration parameter and obtained the best-fit values using $\chi^{2}$ minimization technique with available observational data. They also analyzed the evolution of the jerk parameter for the considered parametrized model. \cite{gadbail2022parametrization} have explored a specific parametrization of deceleration parameter in the context of $f(Q)$ gravity theory and constrained the model parameter by using Bayesian analysis with observational data. Much recently, several theoretical models have been developed to analyze the entire evolutionary history of the Universe through parametrization of $q(z)$ as a function of scale factor $(a(t))$ or time $(t)$ or redshift in \cite{capozziello1,capozziello2,capozziello5,Capozziello3,Capozziello4,Orlando1,Orlando2,Orlando3,Orlando4,Orlando5}.\\\\
Motivated by the above ideas, in this paper we present a parametrization of the deceleration parameter with five unknown parameters. This work mainly focuses on constraining the crucial cosmological parameters of the model using 17 uncorrelated baryon acoustic oscillation (BAO) measurements from the latest galaxy surveys in the redshift range \( z \in [0.106, 2.33] \) \cite{alam2015eleventh}. BAO are subtle fluctuations in the distribution of matter in the Universe that originated from pressure waves in the early Universe \cite{eisenstein2005detection}. These waves traveled through the dense, hot plasma of the early Universe, leaving a characteristic scale imprinted on the distribution of matter when the Universe became transparent to radiation. BAO manifests as a preferred separation scale between galaxies, making them a powerful cosmological tool for studying the expansion history of the Universe. The importance of BAO lies in their ability to serve as standard rulers for measuring cosmic distances at different epochs. By precisely measuring the scale of BAO in large-scale galaxy surveys, one can infer the expansion rate of the universe at different redshifts. This provides crucial constraints on the cosmic expansion history and the nature of dark energy, leading to insights into the underlying cosmological model. The Sound Horizon (\(r_{d}\)), which represents the distance that sound waves could travel in the early Universe before they became transparent, is a fundamental component in BAO measurements \cite{eisenstein1998baryonic}. By accurately determining the Sound Horizon scale, researchers can calibrate the cosmic distance scale, enabling precise measurements of cosmological parameters such as the Hubble parameter and the matter density of the Universe. With the BAO dataset, we also utilize late-time datasets, including Cosmic Chronometers \cite{moresco20166}, Type Ia Supernovae \cite{smith2020first}, Gamma-Ray Bursts \cite{firmani2005new}, and Compact Radio Quasars \cite{quasers}, to extract the best-fit values of cosmological parameters, particularly concerning the Sound Horizon (\(r_{d}\)) and the Hubble parameter (\(H_{0}\)). Late-time datasets contain observational data from relatively recent epochs in cosmic history, allowing us to probe the Universe's expansion in its more recent stages. This late-time data is crucial for constraining cosmological models and understanding the current state of the Universe's evolution, complementing information obtained from early-time observations such as the cosmic microwave background. Therefore, the importance of late-time data lies in its ability to provide vital constraints on the cosmological parameters governing the late-time evolution of the Universe, enhancing our understanding of its fundamental properties. A significant issue in modern cosmology is the Hubble Constant tension \cite{dainotti2021hubble,dainotti2022evolution,montani2024slow,di2021realm,vagnozzi2023seven,vagnozzi2020new}, referring to the discrepancy between \(H_0\) values from early-time measurements, like the cosmic microwave background (CMB), and late-time measurements, such as Cepheid variables and Type Ia supernovae. Planck satellite data suggest a lower \(H_0\) (67-68 km/s/Mpc), while late-time observations indicate a higher value (73-74 km/s/Mpc). This tension suggests potential systematic errors or new physics beyond the standard cosmological model, and resolving it is crucial for understanding the Universe's expansion and underlying physics. Recent advancements in cosmological research have benefited from novel methodologies and refined statistical analyses. Dainotti et al. (2024) introduced a new binning method for Quasar samples, highlighting their potential as standard candles for measuring cosmological distances \cite{NewQuasars1,NewQuasars2}. Their work on Supernovae Ia reduced Hubble constant uncertainties by up to 43\% \cite{NewQuasars3,NewQuasars4}, building on earlier studies that achieved a 35\% reduction through improved analysis of multiple datasets \cite{NewQuasars5}. Quasars have been validated as standard candles up to $z = 7.5$ with precision comparable to Supernovae Ia \cite{NewQuasars6}. The GRB fundamental plane correlation has also been identified as a reliable cosmological tool \cite{NewQuasars7}. A multi-pronged approach, including binned analyses of GRBs, Supernovae Ia, and baryon acoustic oscillations, reinforces the robustness of these methods \cite{NewQuasars8}. Exploration of the X-Ray Fundamental Plane of GRBs, Kilonovae, and SNe Ib/c associated with GRBs has opened new avenues for understanding their relationships and use in cosmology \cite{NewQuasars9,NewQuasars10}. These comprehensive studies enhance our ability to measure cosmic distances and understand the universe's expansion with greater precision.\\\\
The paper is structured as follows: Section (\ref{sec2}) introduces the basic equations of FLRW Universe. Section (\ref{sec3}) we introduces a new five-parameter parametrization of DP. Thereafter, we find the Hubble function corresponding to the parametrization. In Section (\ref{sec4}), the methodologies and datasets have been explained. Section (\ref{sec5}) compares the model's predictions with observational data. Section (\ref{sec6}) gives a detailed description about the kinematic cosmographic parameters such as the deceleration and jerk parameters. Section (\ref{sec7}) and (\ref{sec8}) discuss about the statefinder and $O_{m}$ diagnostics. Section (\ref{sec9}) and Section (\ref{sec10}) discuss the results and conclusions respectively.

\section{Fundamental Equations of the FLRW Model} \label{sec2}
In this section, we delve into the foundational equations of the spatially flat, homogeneous, and isotropic FLRW (Friedmann-Lema\^itre-Robertson-Walker (FLRW)
) universe. This model serves as a cornerstone in our exploration of the cosmos. We embark on our journey by embracing the line element that characterizes the very fabric of space and time:
\begin{equation}
ds^2 = -dt^2 + a^2(t)\left[dr^2 + r^2\left(d\theta^2 + \sin^2 \theta
d\phi^2 \right)\right],
\end{equation}
Here, $a(t)$ unfurls itself as the scale factor, a pivotal entity shaping the evolution of our cosmic theater. Within this cosmic arena, the energy-momentum tensor of the fluid orchestrates the intricate dance of matter and energy. It assumes the form:
\begin{equation}
T_{\mu\nu} = (\rho + p)u_\mu u_\nu + p g_{\mu\nu},
\end{equation}
At its core, $\rho$ and $p$ personify the energy density and pressure density of this cosmic fluid. The 4-velocity $u^\mu = \frac{dx^\mu}{ds}$ weaves the narrative of motion through the cosmos, steadfastly adhering to the profound relationship $u^\mu u_\mu = -1$. The FLRW Universe beckons us to uncover its secrets encoded in the Friedmann equations, sacred inscriptions within Einstein's gravitational framework.\\\\
Friedmann Equation 1:\\
\begin{equation}\label{eq1}
H^2 = \frac{8\pi G}{3}~\rho,
\end{equation}
Friedmann Equation 2:\\
\begin{equation}\label{eq2}
\dot{H} = -4\pi G(\rho + p),
\end{equation}
Here, $H = \frac{\dot{a}}{a}$ emerges as the herald of cosmic expansion, and the temporal derivative denoted by the overhead dot gracefully embodies the cosmic passage of time, $t$. In the realm of these equations, we discern the harmonious interplay between energy densities, pressures, and the cosmic fabric itself, unveiling the essence of our FLRW cosmos. Delving into the intricate tapestry of the cosmos, we find its very essence woven with fluid matter, manifesting as a symphony of energy density $\rho$ and pressure $p$. This cosmic ensemble adheres to a sacred edict, the energy conservation equation, resonating as:

\begin{equation}\label{Cons}
\dot{\rho} + 3H(\rho + p) = 0,
\end{equation}
Here, $H$ is the celestial conductor, the Hubble parameter orchestrating the cosmic rhythm. Our odyssey into the cosmic composition unveils a triad of cosmic constituents: radiation, dark matter (DM), and enigmatic dark energy (DE). Their presence shapes the grand narrative of our universe, with densities and pressures woven into the cosmic fabric: $\rho = \rho_{r} + \rho_{m} + \rho_{d}$ and $p = p_{r} + p_{m} + p_{d}$. Peering into their individual realms, we invoke distinct conservation equations:\\\\
Radiation:\\
\begin{equation}\label{rad}
\dot{\rho}{r} + 3H(\rho{r} + p_{r}) = 0,
\end{equation}
Dark Matter (DM):\\
\begin{equation}\label{DM}
\dot{\rho}{m} + 3H(\rho{m} + p_{m}) = 0,
\end{equation}
Dark Energy (DE):\\
\begin{equation}\label{DE}
\dot{\rho}{d} + 3H(\rho{d} + p_{d}) = 0,
\end{equation}
The radiant dance of radiation is characterized by $p_{r} = \frac{1}{3}\rho_{r}$, yielding the mesmerizing revelation $\rho_{r} = \rho_{r0}a^{-4}$, where $a$ gracefully denotes the cosmic scale factor. Meanwhile, the enigmatic dark matter, a realm of negligible pressure ($p_{m} = 0$), weaves its destiny as $\rho_{m} = \rho_{m0}a^{-3}$. In this symphony of cosmic evolution, these equations breathe life into the dynamic interplay of cosmic components, echoing through the vast cosmic expansion. In our cosmic odyssey, the deceleration parameter, a compass guiding the cosmic voyage, unfurls before us:
\begin{eqnarray}
q = -1 - \frac{\dot{H}}{H^2}
\end{eqnarray}
Venturing further, we unravel the narrative of the enigmatic dark energy. This enigma unfolds as:
\begin{eqnarray}
q_{d} = -1 - \frac{\dot{H_{d}}}{H_{d}^2}
\end{eqnarray}
Here, $H_{d}$ takes on the mantle of the cosmic maestro, conducting the expansion of the dark energy domain. A cosmic duet of equations, \eqref{eq1} and \eqref{eq2}, resonate harmoniously, revealing cosmic truths:
\begin{equation}\label{F11}
H_{d}^2 = \frac{8\pi G}{3}~\rho_{d}
\end{equation}
\begin{equation}\label{F22}
\dot{H}{d} = -4\pi G(\rho{d}+p_{d})
\end{equation}
Stepping onto the cosmic stage with these equations, the ethereal dance of fluid energy density takes center stage:

\begin{equation} \label{rho}
\rho_{d} = \rho_{d0}~ e^{\int \frac{2(1+q_{d})}{1+z}dz}
\end{equation}
A symphony of symbols and cosmic energies converge, where $\rho_{d0}$ embodies the present density parameter, and $z$ portrays the redshift parameter, narrated as $1+z=\frac{1}{a}$ (with $a_{0}=1$ in the present moment). In the cosmic chronicle, celestial constants emerge as cosmic constellations. Casting our gaze upon the cosmic canvas, we define the cosmic tapestry:
$\Omega_{r0}=\frac{8\pi G\rho_{r0}}{3H_{0}^{2}}$,
$\Omega_{m0}=\frac{8\pi G\rho_{m0}}{3H_{0}^{2}}$,
$\Omega_{d0}=\frac{8\pi G\rho_{d0}}{3H_{0}^{2}}$.
From this cosmic calculus, the Hubble parameter script is penned:
\begin{equation}\label{H}
H^{2}(z) = H_{0}^{2}\left[\Omega_{r0}(1+z)^{4} + \Omega_{m0}(1+z)^{3}
+ \Omega_{d0}~ e^{\int \frac{2(1+q_{d})}{1+z}dz}\right]
\end{equation}
In this cosmic sonnet, $\Omega_{d0}$ emerges as the sentinel of cosmic balance, whispering the cosmic tale of unity: $\Omega_{d0}=1-\Omega_{r0}-\Omega_{m0}$.
\section{Parametrization of the deceleration parameter.} \label{sec3}
In our pursuit of deciphering the cosmos, we turn our gaze toward the elusive parameterized deceleration. Its simplest embodiment takes form as follows:
\begin{equation}
q(z) = q_0 + q_1 \mathcal{X}(z)
\end{equation}
Within this cosmic equation, $q_0$ and $q_1$ emerge as steadfast cosmic companions, while $\mathcal{X}(z)$, a function of redshift $z$, guides us through the cosmic terrain. Various cosmic riddles beckon, each demanding a distinct form for $\mathcal{X}(z)$. Yet, previous attempts often left cosmic enigmas unsolved, hinting at the need for a parametrization that truly captures the essence of cosmic evolution.responding Hubble parameter in terms of redshift $z$.
\subsection{(Chaudhary-Debnath-Mustafa-Maurya-Atamurotov) type parametrization}
In our exploration of the cosmos, we introduce a new paradigm that represents a harmonious synthesis of cosmic phenomena.
\begin{equation}\label{New}
q_{d}(z) = q_0 + \frac{\alpha + (1+z)^{\beta}}{q_{1} + q_{2}(1+z)^{\beta}}
\end{equation}
Within this cosmic symphony, $q_{0}$, $q_{1}$, $q_{2}$, $\alpha$ and $\beta$ take on the role of cosmic constants, conducting the cosmic dance. This newfound paradigm ripples through the cosmic tapestry, influencing the very fabric of cosmic energy density:
\begin{equation}\label{rho New}
\rho_d = \rho_{d0}~(1+z)^{3(1+q_0+\frac{\alpha}{q_{1}})}[q_{1}+q_{2}(1+z)^{\beta}]^{\frac{3(q_{1}-\alpha q_{2})}{\beta q_{1}q_{2}}}
\end{equation}
This cosmic crescendo echoes through the cosmic chronicle, reshaping the Hubble parameter:
\begin{widetext}
\begin{equation}
H^{2}(z) = H_{0}^{2}\left[(1+z)^{4}\,\Omega_{r0}+(1+z)^{3}\,\Omega_{m0}
+ (1-\Omega_{r0}-\Omega_{m0})
~(1+z)^{3(1+q_{0}+\frac{\alpha}{q_{1}})}[q_{1}+q_{2}(1+z)^{\beta}]^{\frac{3(q_{1}-\alpha q_{2})}{\beta q_{1}q_{2}}}\right]
\end{equation}
\end{widetext}
In this cosmic symphony, $\Omega_{r0}$ and $\Omega_{m0}$ compose the cosmic orchestra, weaving the cosmic melody that resonates across the cosmos. As we embark on this cosmic voyage, our new model of deceleration parameter charts a course toward understanding the enigmatic cosmic fabric.
\section{Methodology}\label{sec4}
\begin{figure*}
\centering
\includegraphics[scale=0.35]{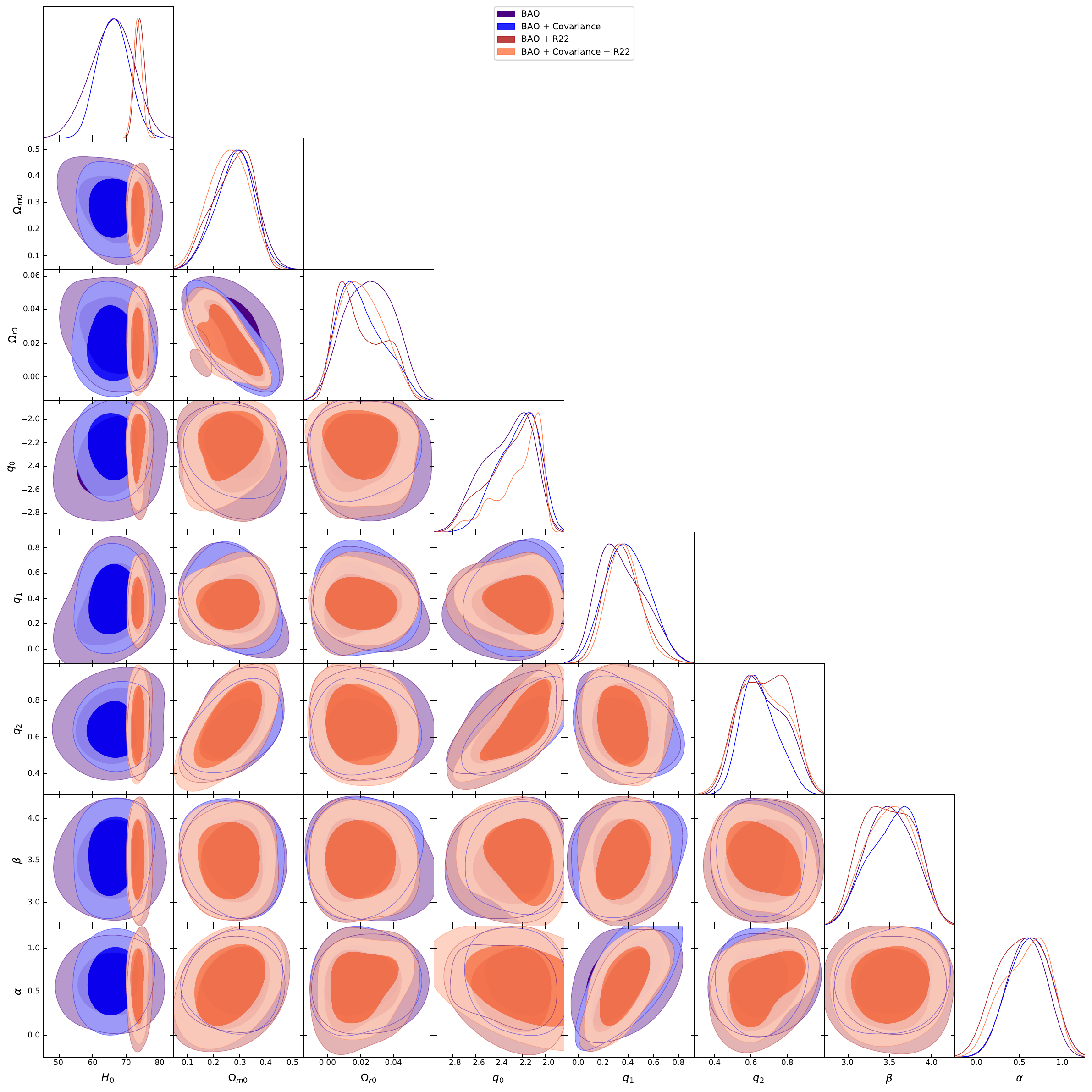}
\caption{The figures illustrate the constraints of the posterior distributions for Model under two scenarios: one with and one without a randomly generated test covariance matrix consisting of twelve components. It is observed that the distribution corresponding to the covariance matrix containing null and twelve components is nearly insignificant, appearing almost indistinguishable from the dataset without any correlations.}\label{fig_1}
\end{figure*}

\begin{figure*}
\centering
\includegraphics[scale=0.35]{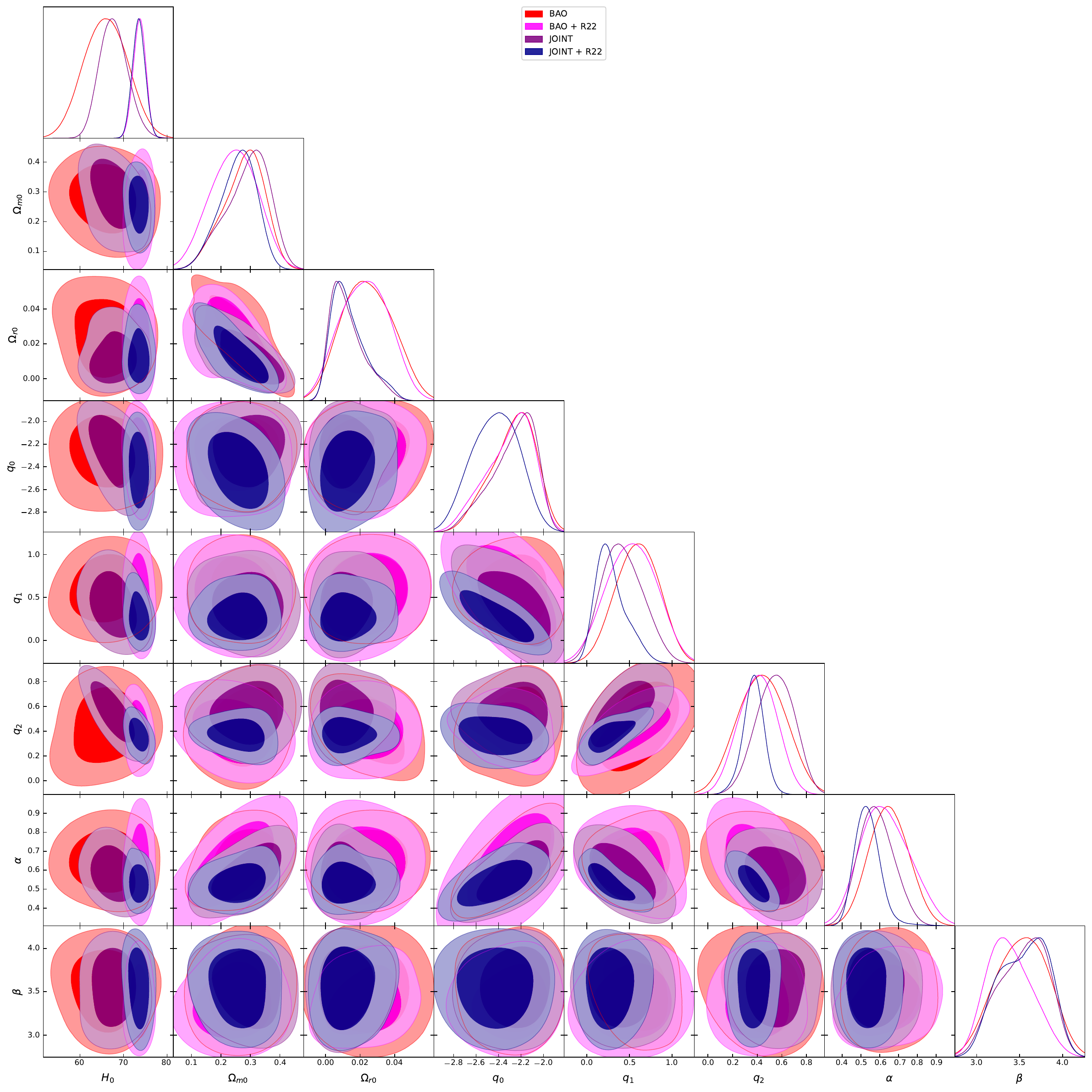}
\caption{The figure presents the parameter constraints obtained from various observational data measurements within the framework of the model, showcasing both 1$\sigma$ and 2$\sigma$ confidence intervals.}\label{fig_2}
\end{figure*}

\begin{figure}
\centering
\includegraphics[scale=0.35]{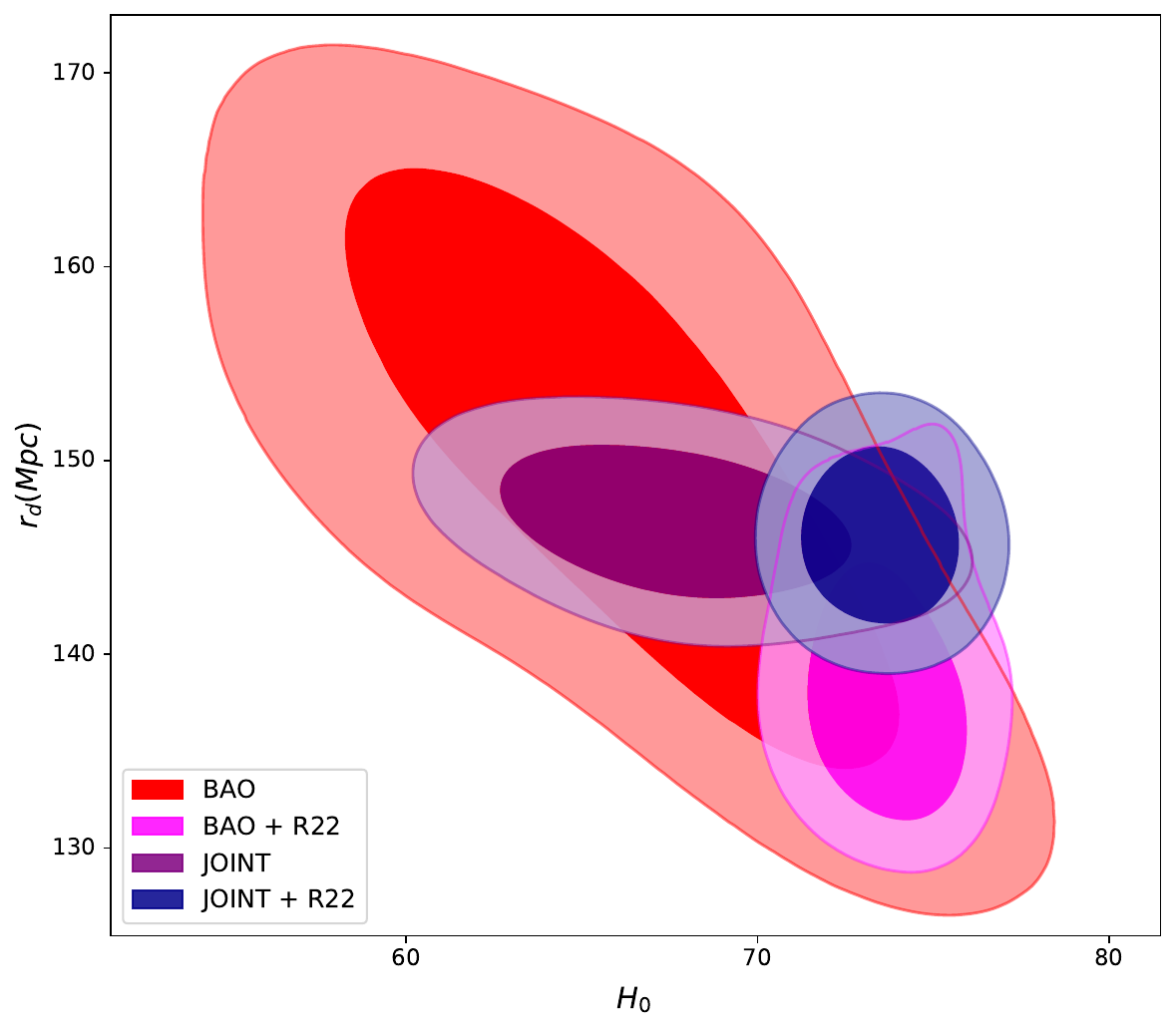}
\caption{The figure presents the parameter constraints obtained from various observational data measurements within the framework of the model, showcasing both 1$\sigma$ and 2$\sigma$ confidence intervals.}\label{fig_3}
\end{figure}

In our investigation, we meticulously curated a subset of the latest Baryon Acoustic Oscillation (BAO) measurements derived from diverse galaxy survey experiments. The primary contributors to this subset include data from the Sloan Digital Sky Survey (SDSS) \cite{42BAO,43BAO,44BAO,45BAO,46BAO,47BAO}, along with contributions from the Dark Energy Survey (DES) \cite{48BAO}, the Dark Energy Camera Legacy Survey (DECaLS) \cite{49BAO}, and 6dFGS BAO \cite{50BAO}. To ensure our analysis adequately captures inherent correlations within the dataset, we recognize the potential for interdependencies among selected data points. Despite our intentional curation to mitigate highly correlated points, addressing potential correlations within our chosen data points remains crucial. Estimating systematic errors necessitates employing mock data derived from N-body simulations to accurately determine covariance matrices. Given the diverse set of measurements from various observational surveys, precisely determining covariance matrices between them proves challenging. To overcome this challenge, we adopt the covariance analysis approach proposed in \cite{38BAO}, where $C_{ii} = \sigma_{i}^{2}$. To simulate correlations within our selected subsample, we introduce non-diagonal elements into the covariance matrix while preserving symmetry. Incorporating non-negative correlations involves randomly selecting up to twelve pairs of data points and assigning non-diagonal elements with magnitudes of $C_{ij} = 0.5 \sigma_{i} \sigma_{j}$, where $\sigma_i$ and $\sigma_j$ represent the 1$\sigma$ errors of data points $i$ and $j$. The locations of these non-diagonal elements are chosen randomly, allowing us to effectively represent correlations within 65\% of the chosen BAO dataset. Fig~\ref{fig_1} depicts the posterior distributions for our cosmological Model with and without the inclusion of a test random covariance matrix consisting of twelve components. It is evident that the influence of the covariance matrix, ranging from null to twelve components, is minimal, closely resembling that of the uncorrelated dataset. To determine the best-fit values of our cosmological model parameters, we expand the BAO dataset by incorporating thirty uncorrelated Hubble parameter measurements obtained through the cosmic chronometers (CC) method discussed in \cite{CC1,CC2,CC3,CC4}. Additionally, we include the binned Type Ia Supernovae dataset \cite{smith2020first}. Expanding our observational scope further, we introduce 24 binned quasar distance modulus data from \cite{Quasar}, a set of 145 Gamma-Ray Bursts (GRBs) has been formed by combining 50 from the Platinum sample and 95 from the LGRB95 sample, as outlined in \cite{Dainotti:2020jkj} and further detailed in \cite{cao2022gamma}, and the recent Hubble constant measurement (R22) \cite{7BAO} as an additional prior. The new likelihood approach discussed in \cite{bargiacchi2023gamma,dainotti2023stochastic} has been applied to this dataset. Our analysis utilizes a nested sampling approach implemented in the open-source Polychord package \cite{Polychord}, complemented by the GetDist package \cite{Getdist} for presenting results. Figs~\ref{fig_2} depict the \(68\%\) and \(95\%\) confidence levels for key cosmological parameters for the new parametrization of the deceleration parameter. Table~\ref{tab_2} presents the best-fit values of this new parametrization along with the current Hubble constant \(H_{0}\). Fig~\ref{fig_3} illustrates the posterior distribution of the \(r_{d}-H_{0}\) contour plane.
\begin{table*}
\centering
\begin{tabular}{|c|c|c|c|c|c|c|}
\hline
\multicolumn{7}{|c|}{MCMC Results} \\
\hline\hline
Model & Parameters & Priors & BAO & BA0 + R22 & JOINT & JOINT + R22 \\[1ex]
\hline
& $H_0$ & [50,100] & $69.089209_{\pm 4.396874}^{\pm 6.252595}$ & $73.906596_{\pm 1.357697}^{\pm 2.788642}$ & $69.854848_{\pm 1.259100}^{\pm 2.386935}$ & $71.616475_{\pm 1.003940}^{\pm 1.936192}$ \\[1ex]
$\Lambda$CDM Model &$\Omega_{m0}$ &[0.,1.]  & $0.256796_{\pm 0.025772}^{\pm 0.069423}$ & $0.254859_{\pm 0.026912}^{\pm 0.068716}$ & $0.268654_{\pm 0.012822}^{\pm 0.028134}$ & $0.264407_{\pm 0.013179}^{\pm 0.030224}$ \\[1ex]
& $r_d$ (Mpc) & [100,200] & $149.504236_{\pm 10.037166}^{\pm 15.212831}$ & $139.447508_{\pm 2.914639}^{\pm 5.883881}$ & $146.543556_{\pm 2.598566}^{\pm 5.101856}$ & $143.299915_{\pm 2.218062}^{\pm 4.358875}$ \\[1ex]
& $r_{d}/r_{fid}$ & [0.9,1.1] & $1.008382_{\pm 0.066344}^{\pm 0.101188}$ & $0.940599_{\pm 0.020806}^{\pm 0.037308}$ & $0.990266_{\pm 0.019826}^{\pm 0.035810}$ & $0.967284_{\pm 0.015101}^{\pm 0.030444}$ \\
\hline
& $H_{0}$ & [50.,80.] & $65.914970_{\pm 4.964165}^{\pm 8.650553}$ & $73.670021_{\pm 1.441869}^{\pm 2.906876}$ & $67.706623_{\pm 3.183443}^{\pm 5.372325}$ & $73.506493_{\pm 1.484860}^{\pm 2.686997}$ \\[1ex]
& $\Omega_{m0}$ & [0.,1.] & $0.278189_{\pm 0.075665}^{\pm 0.147359}$ & $0.244292_{\pm 0.095621}^{\pm 0.135582}$ & $0.290668_{\pm 0.088132}^{\pm 0.156972}$ & $0.258608_{\pm 0.066012}^{\pm 0.133442}$ \\[1ex]
& $\Omega_{r0}$ & [0.,0.1] & $0.024186_{\pm 0.015378}^{\pm 0.022731}$ & $0.022756_{\pm 0.016716}^{\pm 0.021626}$ & $0.012523_{\pm 0.009592}^{\pm 0.012239}$ & $0.013411_{\pm 0.009959}^{\pm 0.012799}$ \\[1ex]
& $q_{0}$ & [-2.8,-2.] & $-2.271204_{\pm 0.211664}^{\pm 0.415293}$ & $-2.297193_{\pm 0.218924}^{\pm 0.463253}$ & $-2.265561_{\pm 0.218810}^{\pm 0.451835}$ & $-2.425894_{\pm 0.230311}^{\pm 0.355341}$ \\[1ex]
New Model & $q_{1}$ & [0.,0.8] & $0.591868_{\pm 0.244407}^{\pm 0.426907}$ & $0.524783_{\pm 0.313358}^{\pm 0.478415}$ & $0.417265_{\pm 0.248408}^{\pm 0.389058}$ & $0.279974_{\pm 0.173272}^{\pm 0.261155}$ \\[1ex]
& $q_{2}$ & [0.,0.9] & $0.435534_{\pm 0.209352}^{\pm 0.342693}$ & $0.404773_{\pm 0.164544}^{\pm 0.270470}$ & $0.545593_{\pm 0.141437}^{\pm 0.309750}$ & $0.369376_{\pm 0.071240}^{\pm 0.184829}$ \\[1ex]
& $\alpha$ & [0.,1.] & $0.647908_{\pm 0.104986}^{\pm 0.186600}$ & $0.630145_{\pm 0.119284}^{\pm 0.217269}$ & $0.581334_{\pm 0.090477}^{\pm 0.158326}$ & $0.533942_{\pm 0.065614}^{\pm 0.106099}$ \\[1ex]
& $\beta$ & [3.,4.] & $3.529518_{\pm 0.344931}^{\pm 0.495478}$ & $3.385145_{\pm 0.283510}^{\pm 0.363824}$ & $3.547634_{\pm 0.353282}^{\pm 0.523800}$ & $3.553258_{\pm 0.329610}^{\pm 0.512899}$ \\[1ex]
& $r_{d}$ (Mpc) & [100,200] & $150.260293_{\pm 10.808205}^{\pm 15.616586}$ & $138.298819_{\pm 4.110937}^{\pm 5.902072}$ & $146.884404_{\pm 2.276929}^{\pm 4.566931}$ & $146.081377_{\pm 2.809924}^{\pm 4.971972}$ \\[1ex]
& $r_{d}/r_{fid}$ & [0.9,1.1] & $1.012950_{\pm 0.078681}^{\pm 0.101536}$ & $0.934839_{\pm 0.026944}^{\pm 0.033286}$ & $0.991616_{\pm 0.017502}^{\pm 0.038852}$ & $0.986487_{\pm 0.019074}^{\pm 0.031595}$ \\[1ex]
\hline
\end{tabular}
\caption{The table presents constraints on cosmological parameters for the $\Lambda$CDM and New Model at a 95\% confidence level.}\label{tab_2}
\end{table*}

\section{Observational, and theoretical comparisons with Hubble Measurements}\label{sec5}
After determining the values of the free parameters in our cosmological model, it is crucial to compare the model's predictions with observational data and the well-established $\Lambda$CDM paradigm. This step allows us to assess the model's viability and its ability to explain observed phenomena in the Universe. By contrasting the model predictions with observational data, we can evaluate the agreement and identify any discrepancies or deviations from the standard $\Lambda$CDM framework. This analysis plays a crucial role in validating or refining the proposed cosmological model and advancing our understanding of the underlying physical processes driving the evolution of the Universe.
\subsection{Contrasting with Hubble Measurements} 
\begin{figure}
\centering
\includegraphics[scale=0.4]{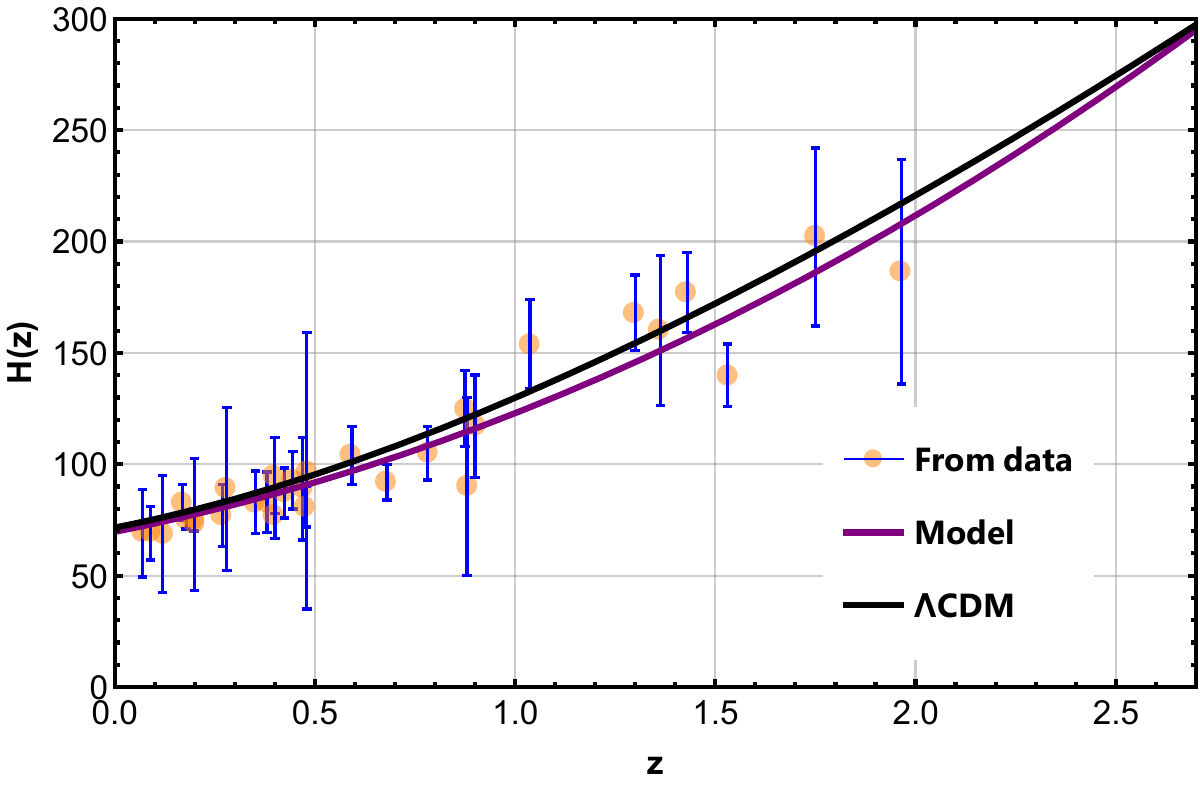}
\caption{Comparative analysis of our model ( Purple line ) with 30 Cosmic Chronometers datasets ( orange dots ) and $\Lambda$CDM model ( black line )}\label{fig_4}
\end{figure}

We assess the compatibility of our model with observational data by comparing it to the 30 measurements of Cosmic Chronometers datasets, represented by orange dots accompanied by error bars in the purple line shown in Fig~\ref{fig_4}. For reference, we also include the well-established $\Lambda$CDM model in black line with $\Omega_{\mathrm{m0}}=$ 0.3 and $\Omega_\Lambda =$ 0.7. The comparative analysis provides essential insights into how well our model performs in describing the observed Hubble data. The figure shows a close alignment between our model's predictions and the orange data points, indicating that our model successfully captures the features and trends present in the dataset. This alignment demonstrates that our model can reproduce the expansion history of the universe as inferred from the Hubble data. The agreement between our model and the observed cosmic evolution supports its validity and credibility.
\subsection{Relative differences between Model and $\Lambda$CDM.}
\begin{figure}
\centering
\includegraphics[scale=0.4]{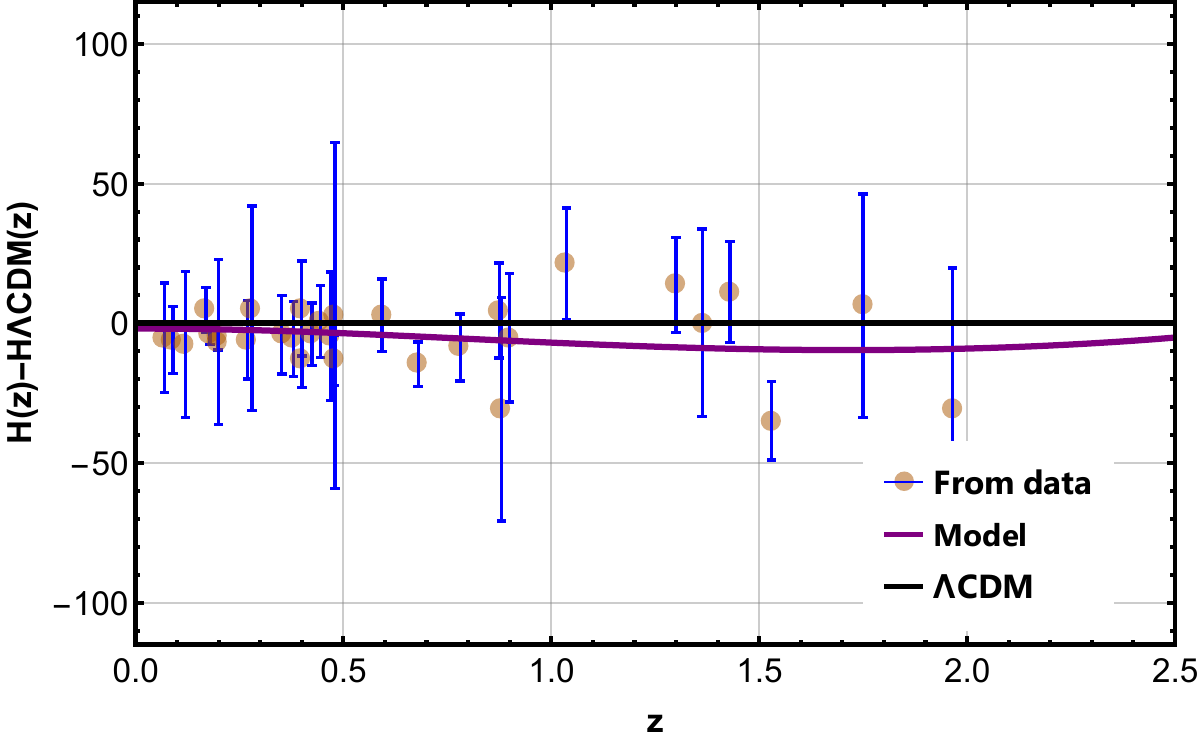}
\caption{Comparative analysis of our model ( Purple line ) with 30 Cosmic Chronometers datasets ( orange dots ) and $\Lambda$CDM model ( black line )}\label{fig_5}
\end{figure}

The relative distinctions between our model and the $\Lambda$CDM model, we present Fig~\ref{fig_5}. The plot illustrates the behaviors of both models for redshifts $z<1$, where they exhibit comparable dynamics. However, as the redshift surpasses unity, slight discrepancies emerge between the two models. These differences are particularly noticeable in the higher redshift range. The deviations between our model and the $\Lambda$CDM model gradually diminish beyond a critical redshift of approximately $z\approx 0.5$. This convergence at lower redshifts indicates that both models tend to align in their predictions as we explore earlier cosmic epochs. The close resemblance between our model and the $\Lambda$CDM model for $z<1$ suggests that both models adequately capture the observed cosmic dynamics within this range. However, the observed discrepancies for higher redshifts indicate the presence of unique features or alternative mechanisms in our model, potentially influencing the Universe's evolution at those epochs.
\section{Cosmographic Parameters} \label{sec6}
Cosmographic parameters are mathematical quantities used to describe the behavior of the universe and its expansion. These parameters, such as the deceleration and jerk parameter, are derived from the Taylor series expansion of the scale factor and its derivatives.They give important insights into the dynamics and history of the cosmos, allowing us to investigate the acceleration or deceleration of its expansion as well as higher-order properties \cite{visser2004jerk}. Researchers can acquire a better grasp of the underlying physics and test alternative hypotheses of dark energy by analysing and comparing these cosmographic features \cite{bamba2012dark}.
\subsection{The deceleration parameter}
In cosmology, the deceleration parameter is a dimensionless quantity used to determine the pace of expansion of the universe. It informs us about the accelerated or decelerated stages of cosmic expansion. It is defined mathematically as  \cite{bolotin2015cosmology}.
\begin{equation}
q = -\frac{a\ddot{a}}{\dot{a}^2}.
\end{equation}
The DP influences the behavior of the Hubble parameter, which characterizes the rate at which the universe is expanding \cite{capozziello2022model}. Depending on the sign of the deceleration parameter, the Hubble parameter can either increase or decrease. Positive values of $q$ indicate an accelerated expansion of the universe, while negative values correspond to a decelerated expansion. Understanding and estimating the deceleration parameter is crucial for exploring the range of possible values denoted by $q_{0}$ through observational analysis. One way to probe the deceleration parameter is by studying the apparent brightness and redshift of supernovae in distant galaxies \cite{john2004cosmographic}. By examining these observational data, researchers can gain insights into the behavior of the deceleration parameter and its impact on the universe's expansion. Although estimating the deceleration parameter may seem challenging, recent findings strongly support models that suggest an accelerating universe \cite{naik2023observational}. These new outcomes have greatly favored the notion of an expanding universe with an accelerated rate of expansion. In cosmology, there is often a desire to determine a precise value for $q_{0}$, representing the deceleration parameter at the present epoch. However, achieving such precision requires rigorous observational analysis and careful consideration of various astrophysical phenomena \cite{camarena2020local}. Overall, the deceleration parameter provides valuable insights into the dynamics of cosmic expansion. By examining its behavior and exploring its implications through observational studies, scientists can deepen their understanding of the Universe's evolution.
\begin{figure}
\centering
\includegraphics[scale=0.4]{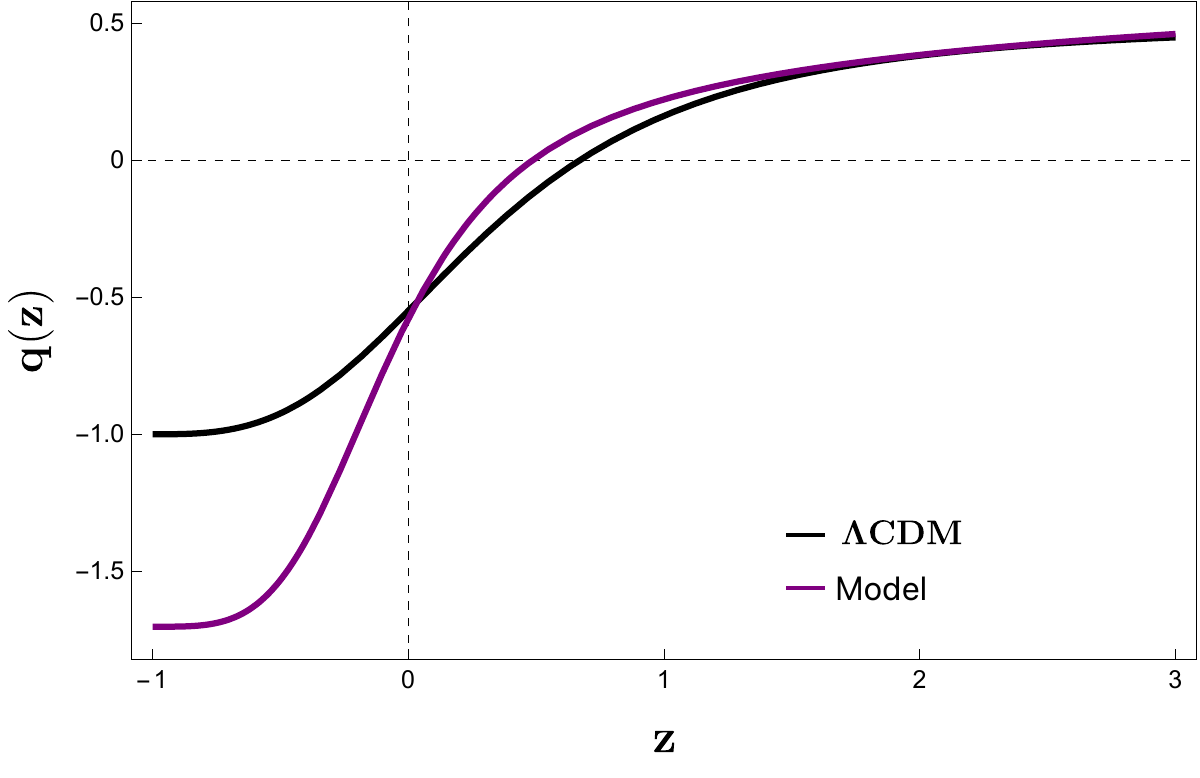}
\caption{This figure visually compares the deceleration parameter between the New Cosmological Model (Purple Line) and the $\Lambda$CDM Model (Black Line).}\label{fig_6}
\end{figure}
\subsection{The jerk parameter}
The dimensionless jerk parameter provides a generalization of the conventional cosmological parameters, such as the scale factor $a(t)$ and the deceleration parameter $q$. It arises from the fourth term in a Taylor series expansion of the scale factor around a reference time $t_0$. The expansion is given by:
\begin{equation}
\begin{aligned}
\frac{a(t)}{a_0} & =1 + (t - t_0)\,H_0 - \frac{(t - t_0)^2}{2}\,H_0^2\,q_0 + \frac{(t - t_0)^3}{6}\,H_0^3\,j_0 \\ &+O\left[(t - t_0)^4\right],
\end{aligned}
\end{equation}
where subscript $0$ denotes the present values of the parameters. This expansion captures the behavior of the scale factor around $t_0$ up to the third derivative. Mathematically, the jerk parameter $j$ is defined as the third derivative of the scale factor with respect to cosmic time, normalized by the ratio of the first derivative of the scale factor to the scale factor itself:
\begin{equation}
j = \frac{1}{a}\frac{d^3a}{d\tau^3}\left(\frac{1}{a}\frac{da}{d\tau}\right)^{-3} = q(2q + 1) + (1 + z)\frac{dq}{dz}.
\end{equation}
The jerk parameter plays a significant role in distinguishing various proposals for dark energy (DE) and their implications for the dynamics of the universe. It helps identify favorable candidates for the physical interpretation of cosmic dynamics amidst the wide range of DE possibilities. By examining specific values of the jerk parameter, one can establish correspondences between DE proposals and standard models of the universe. Models involving a cosmic jerk provide a more pronounced demonstration of transitions between different eras of accelerated expansion. Notably, the flat $\Lambda$CDM model corresponds to a jerk parameter value of $j = 1$.
\begin{figure}
\centering
\includegraphics[scale=0.4]{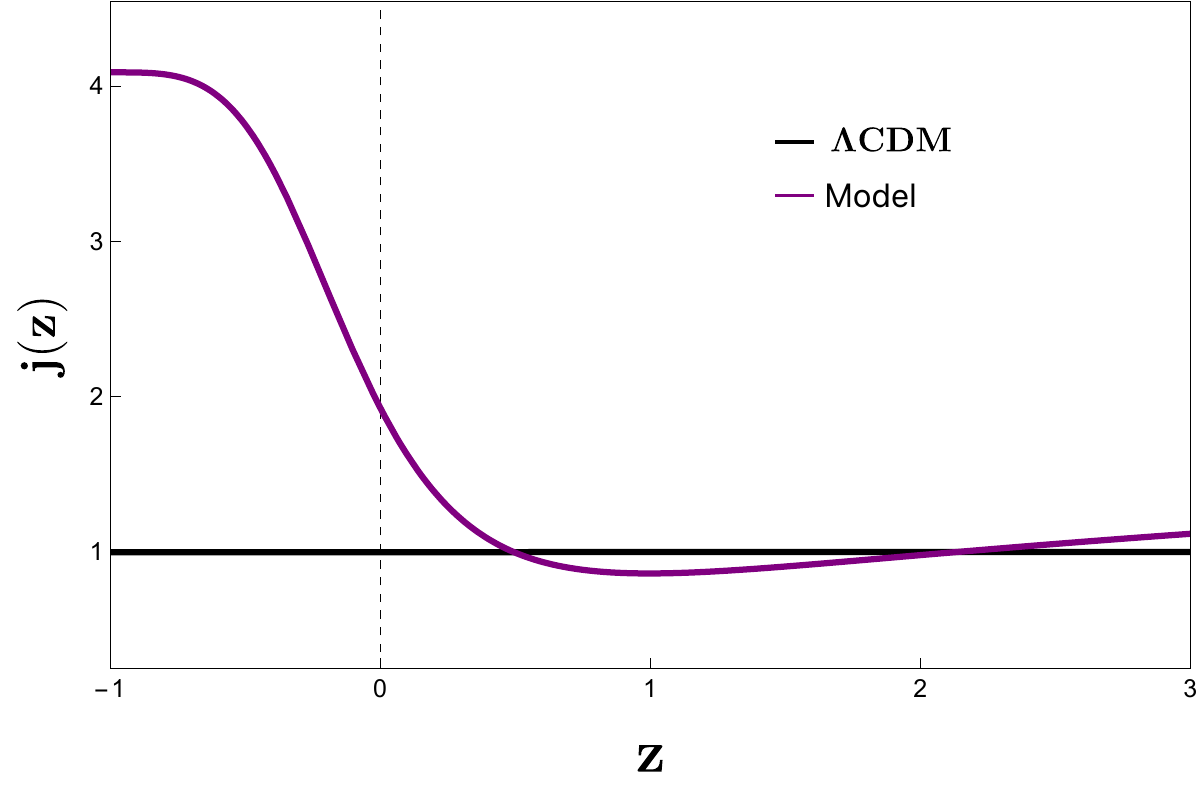}
\caption{This figure visually compares the jerk parameter between the New Cosmological Model (Purple Line) and the $\Lambda$CDM Model (Black Line).}\label{fig_7}
\label{fig6}
\end{figure}
\section{Statefinder Diagnostic} \label{sec7}
The statefinder diagnostic pair $\{r, s\}$ is a widely used tool in the study of dark energy (DE) models, offering insights into their nature based on higher-order derivatives of the scale factor \cite{sahni2003statefinder}. These dimensionless parameters provide a model-independent way to analyze the cosmic properties of DE. The statefinder pair is computed using the following expressions:
\begin{equation}
r=\frac{\dddot{a}}{a H^3}, \quad s=\frac{r-1}{3\left(q-\frac{1}{2}\right)},
\end{equation}
Here, $r$ represents the ratio of the third derivative of the scale factor to the cube of the Hubble parameter, and $s$ is a linear combination of $r$ and the deceleration parameter $q$. Specific values of $r$ and $s$ have well-established interpretations within standard DE models. For example, $\{r,s\}=\{1,0\}$ corresponds to the $\Lambda$CDM model, while $\{r,s\}=\{1,1\}$ corresponds to the standard cold dark matter (SCDM) model in a Friedmann-Lemaître-Robertson-Walker (FLRW) universe. The value range of $(-\infty, \infty)$ indicates the possibility of an Einstein static universe. In the $r-s$ plane, different regions correspond to distinct types of DE models. For instance, $s > 0$ signifies quintessence-like models, whereas $s < 0$ indicates phantom-like models. Deviations from the standard value of $\{r,s\}=\{1,0\}$ can indicate an evolutionary process from phantom-like to quintessence-like behavior. Moreover, specific combinations of the deceleration parameter $q$ and the statefinder parameter $r$ are associated with well-known models. The pair $\{q, r\}=\{-1,1\}$ is linked to the $\Lambda$CDM model, while $\{q, r\}=\{0.5,1\}$ corresponds to the SCDM model.
\begin{figure}
\centering
\includegraphics[scale=0.42]{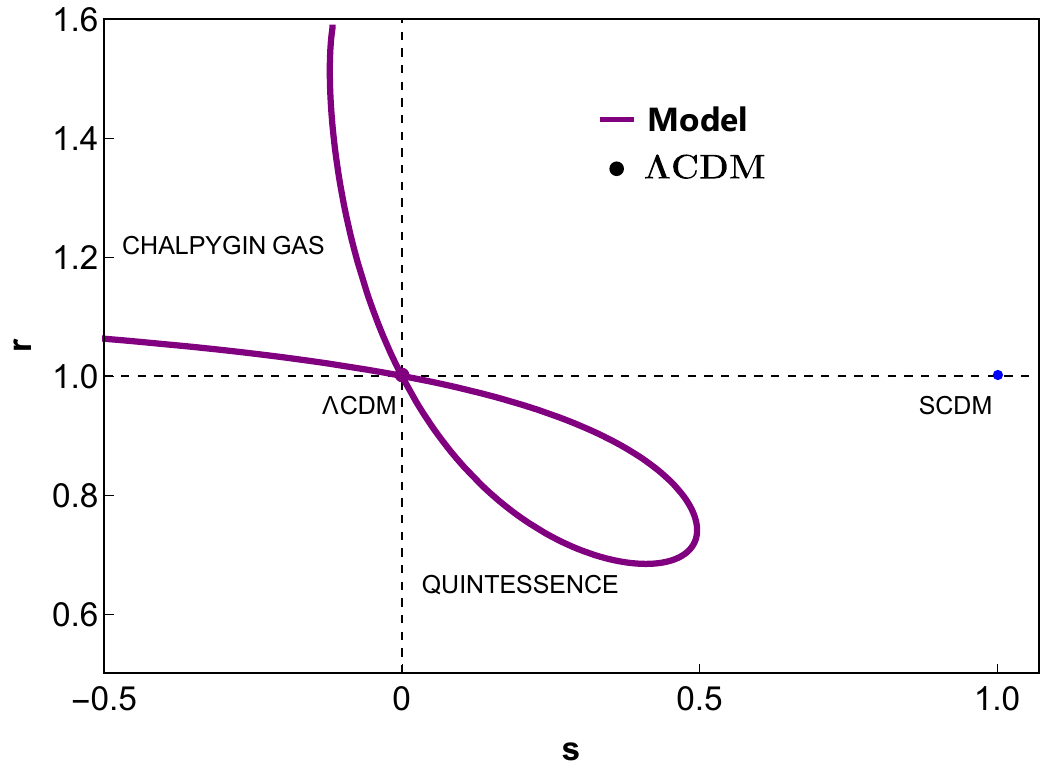}
\caption{Behavior of $\left\{r, s\right\}$ plane}\label{fig_8}
\end{figure}
\begin{figure}
\centering
\includegraphics[scale=0.42]{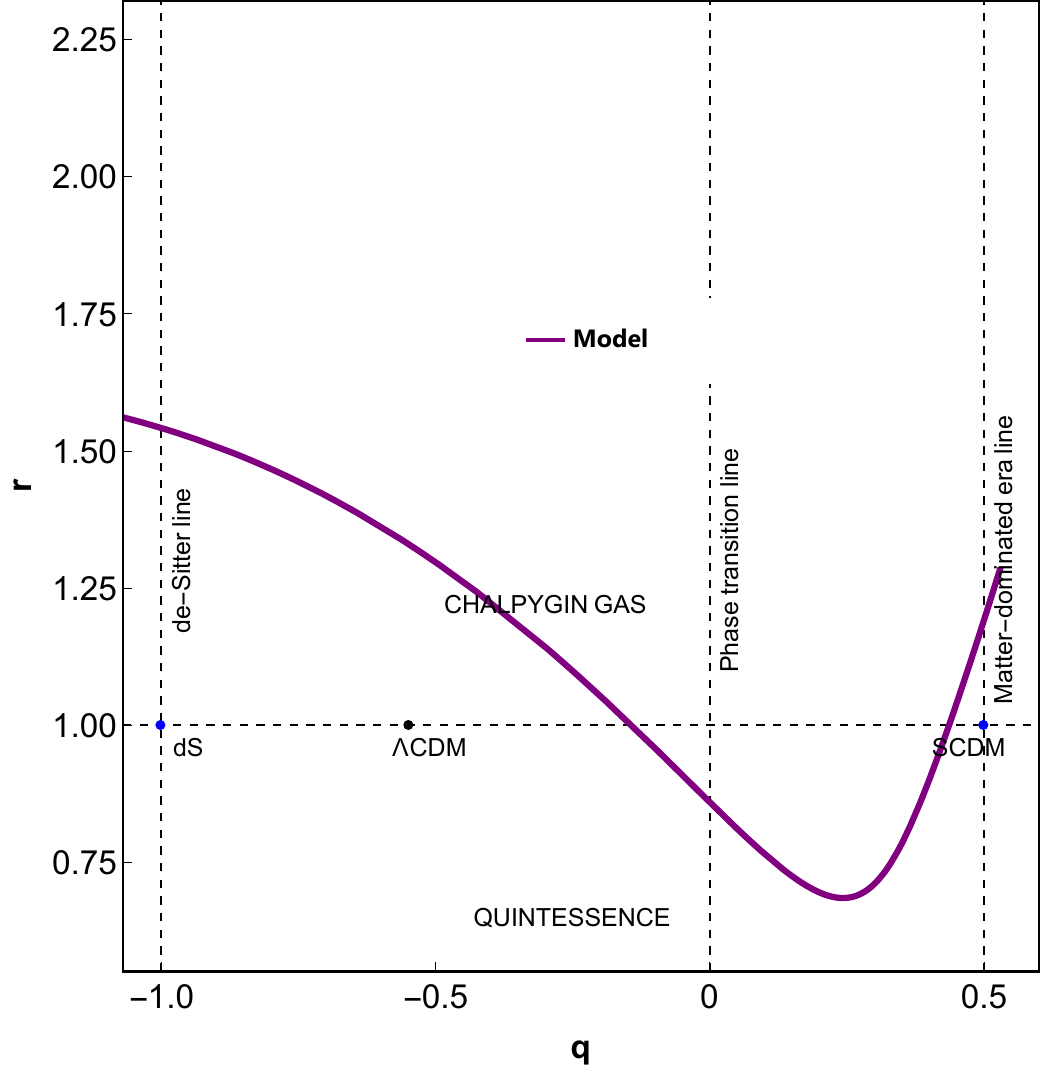}
\caption{Behavior of $\left\{r, q\right\}$ plane}\label{fig_9}
\end{figure}
\section{$O_{m}$ Diagnostic Test}\label{sec8}
In our study, we utilize a powerful diagnostic tool, the $O_{m}$ diagnostic, to investigate Dark Energy (DE) properties. This diagnostic method, introduced in \cite{sahni2008two}, offers a straightforward approach, relying solely on the directly measurable Hubble parameter $H(z)$ obtained from observational data. The $O_{m}$ diagnostic serves as a valuable instrument for distinguishing between different cosmological scenarios, specifically discerning the cosmological constant indicative of a standard $\Lambda$CDM model from a dynamic model associated with a curved $\Lambda$CDM. This discrimination is achieved by comparing the values of $O_{m}$ and $\Omega_{m0}$. When $O_{m}$ equals $\Omega_{m0}$, it indicates consistency with the $\Lambda$CDM model. Conversely, situations where $O_{m}$ exceeds $\Omega_{m0}$ suggest a quintessence scenario, while $O_{m}$ being less than $\Omega_{m0}$ points to a phantom scenario \cite{escamilla2016nonparametric,holsclaw2011nonparametric}. This diagnostic test not only offers a robust approach to understanding Dark Energy but also provides a unique method for distinguishing between various cosmological models. Its reliance on observational measurements, particularly the Hubble parameter, enhances its practicality and reliability in revealing the fundamental dynamics of the Universe.
In a flat Universe, the expression for $O_{m}$ is defined as:
\[ O_{m} = \frac{\Omega_{m0}}{E(z)^{2}} \]
Where $\Omega_{m0}$ represents the present-day matter density parameter, and $E(z)$ is the dimensionless Hubble parameter.
\begin{figure}
\centering
\includegraphics[scale=0.4]{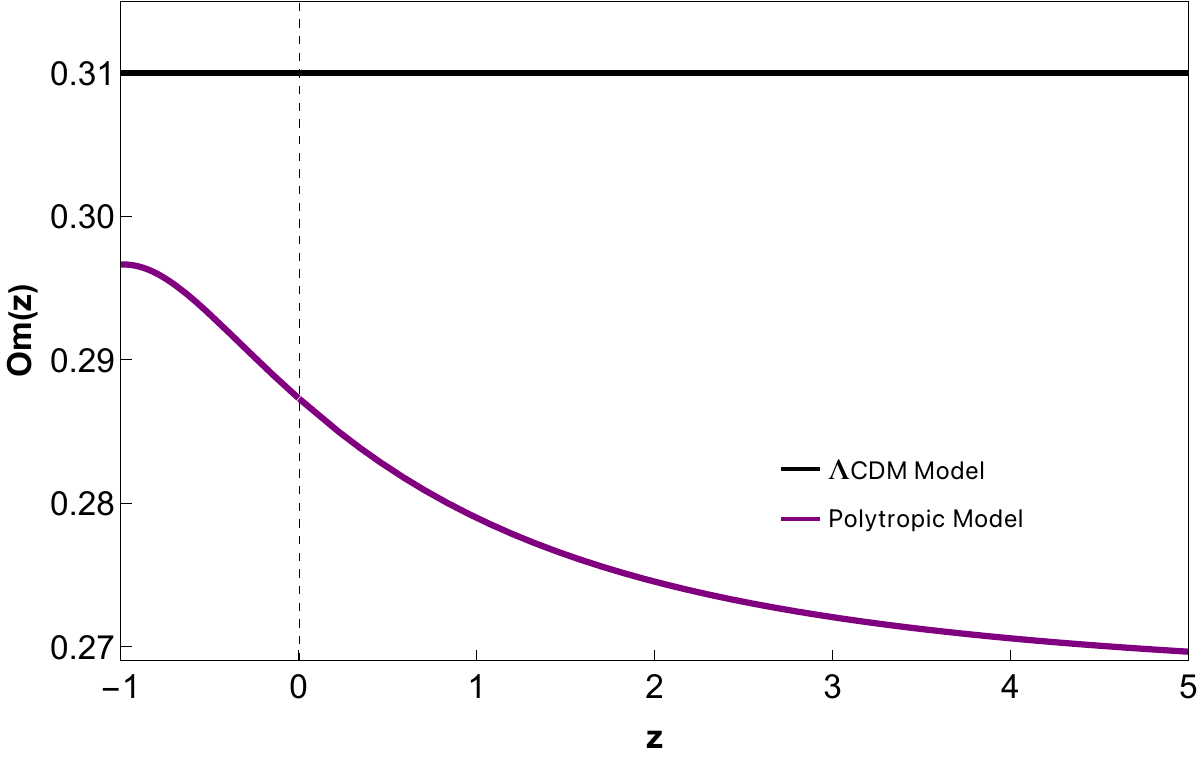}
\caption{Behavior of $Om(z)$ Profile}\label{fig_10}
\end{figure}
\section{Results}\label{sec9}
\paragraph{Cosmological parameters:}
Upon integrating the R22 prior into the Joint dataset, the best-fitting value for \(H_{0}\) is determined as \(73.50 \pm 1.48\), which deviates from the findings in \cite{1BAO} but closely matches measurements from the SNIe sample in \cite{7BAO}. Conversely, without R22 priors in the Joint dataset, the estimated \(H_{0}\) is \(67.70 \pm 3.18\), aligning more closely with the value reported in \cite{1BAO}. Notably, the optimal values obtained for the matter density in the new model, denoted as $\Omega_{m0}$, appear to be lower compared to the values documented in \cite{1BAO} ($\Omega_{m0}=0.315 \pm 0.007$). In the context of Baryon Acoustic Oscillations (BAO), the BAO scale \(r_d\) is characterized by the cosmic sound horizon imprinted during the drag epoch, represented as \(z_d\), which marks the separation of baryons and photons. It is derived from the integral of the ratio of the speed of sound (\(c_s\)) to the Hubble parameter (\(H\)) over the redshift range from \(z_d\) to infinity. The speed of sound, \(c_s\), is determined by \(\frac{1}{\sqrt{3(1+R)}}\), where \(R\) is the ratio of the baryon density perturbation to the photon density perturbation. Observational data from \cite{1BAO} provides \(z_d = 1059.94 \pm 0.30\), and for a $\Lambda$CDM model, \(r_d\) is estimated to be \(147.09 \pm 0.26\) Mpc \cite{1BAO}. In the context of the New Cosmological Model, For the BAO datasets, the resulting \(r_{d}\) is \(150.26 \pm 10.80\) Mpc. However, when exclusively incorporating the R22 prior into the BAO dataset, the sound horizon at the drag epoch is determined to be \(138.29 \pm 4.11\) Mpc. In the case of the Joint dataset, the determined \(r_{d}\) is \(146.88 \pm 2.27\) Mpc, closely aligning with the findings reported in \cite{1BAO}. Furthermore, the integration of the R22 prior into the full dataset results in \(r_{d} = 146.08 \pm 2.80\) Mpc, showing proximity to the findings presented in \cite{1BAO}. These results suggest a reduction in tension in \(r_{d}\).
\\\\ 
\paragraph{deceleration parameter:}
Fig. \ref{fig_6} illustrates the changes in the DP with respect to the cosmic redshift ($z$) in our new cosmological model. Remarkably, our model demonstrates an excellent agreement with the standard $\Lambda$CDM model at redshifts. However, the real excitement lies in the future phase of the Universe's evolution, where our new model showcases a fascinating phenomenon a super accelerated expansion ($q(-1) \approx -1.767$). This stands in stark contrast to the $\Lambda$CDM model, which predicts a de Sitter phase with $q(-1) = -1$. The discovery of this super-accelerated expansion has significant implications, as it suggests that the Universe might experience unprecedented and rapid expansion in the distant future. Indeed, the observed super-accelerated expansion in the new cosmological model could potentially be attributed to dark energy.\\\\
\paragraph{jerk parameter:}
Fig \ref{fig_7} illustrates the evolution of the jerk parameter with cosmic redshift ($z$). At high redshifts, our models exhibit a close agreement with the well-established $\Lambda$CDM model, as the jerk value lies near 1. However, as the redshift decreases, there is a noticeable increase in the value of the jerk parameter for our cosmological model. Specifically, the jerk parameter in our case predicts a value of $j(-1) \approx 4.15$, which is higher than the value predicted by $\Lambda$CDM at $j(-1) \approx 1$. The increase in the jerk parameter value with decreasing redshift suggests a departure from the standard behavior predicted by the $\Lambda$CDM model. In cosmology, the jerk parameter characterizes the rate of change of acceleration with respect to time. A higher jerk parameter indicates that the rate of change of acceleration is increasing more rapidly over time compared to the $\Lambda$CDM model.\\\\
\paragraph{Statefinder Diagnostic}
In Fig \ref{fig_8}, we present the intriguing evolution of the $\{r,s\}$ parameter. Our model depicts an interesting journey through several dark energy domains, offering information on the universe's dynamics. Our model begins in the Chalpygin gas-type dark energy area, which is defined by $r > 1$ and $s < 0$. As the cosmic evolution progresses, the model crosses the fixed $\Lambda$CDM point $\{1,0\}$. Following that, it reaches the Quintessence type dark energy domain, with $\{r,s\}$ parameter values of $r < 1$ and $s > 0$. Subsequently it goes through another transition, returning to the Chalpygin gas-type dark energy region defined by $r > 1$ and $s < 0$ by crossing fixed $\Lambda$CDM point $\{1,0\}$. This back-and-forth movement across the fixed $\Lambda$CDM point highlights the complexity and richness of the dynamics inherent in our proposed model. Fig \ref{fig_9} illustrates the intriguing evolution of the $\{r, q\}$ parameter in our cosmological model. At the outset of cosmic evolution, our model resides in the Chalpygin gas-dominated DE region, characterized by $r > 1$ and $q > 0$. As cosmic time progresses, the model transitions into the Quintessence-dominated DE region, characterized by $r < 1$ and $q < 0$. Subsequently, the model undergoes a phase transition from the Quintessence domain back to the Chalpygin gas DE region.\\\\
\paragraph{$O_{m}$ diagnostic:}
Fig \ref{fig_10} illustrates that in the context of the new cosmological models, the $O_{m}$ value is consistently remains lower than the present matter density parameter \( \Omega_{m0} \) across all redshifts. This observation suggests that the proposed models fall within the phantom region, characterized by an effective EoS for dark energy, \( w_{\text{eff}} \), that is less than -1. The phantom region is associated with exotic forms of dark energy that violate the null energy condition, leading to peculiar and accelerated expansion of the Universe.\\\\
\section{Discussions and Concluding Remarks} \label{sec10}
The present paper introduces a novel cosmological model for investigating the Universe's evolution, employing a fresh parametrization of the deceleration parameter. The model considered a spatially flat, homogeneous, and isotropic Friedmann-Lema\^itre-Robertson-Walker (FLRW) universe filled with radiation, dark matter (DM), and dark energy (DE). They derived the Friedmann equations and the energy conservation equation for the Universe and considered separate conservation equations for radiation, DM, and DE. The new model proposed for the deceleration parameter was given as $q_{d}(z) = q_0 + \frac{\alpha + (1+z)^{\beta}}{q_{1}+q_{2}(1+z)^{\beta}}$, where $q_{0}$, $q_{1}$, $q_{2}$, $\alpha$, and $\beta$ are constants. The author utilizes various datasets, including 30 data points from Hubble parameter measurements obtained through the cosmic chronometers (CC) method, the Pantheon SNeIa dataset comprising 40 data points, 24 binned quasar data, 145 Gamma-Ray Bursts (GRBs), 17 BAO points and recent measurement of the Hubble constant (R22), to constrain crucial cosmological parameters. These parameters include $\Omega_{r0}$, $\Omega_{m0}$, $q_{2}$, $q_{1}$, $q_{0}$, $\alpha$, and $\beta$, in conjunction with the present-day Hubble value ($H_{0}$) and Sound Horizon ($r_{d}$) as a free parameter. We have derived specific values for $H_0$ and $r_d$ in our new cosmological model. Our analysis reveals $H_0 = 67.70 \pm 3.18 \mathrm{km/s/Mpc}$ and $r_d = 146.88 \pm 2.27 \ \mathrm{Mpc}$. An important observation is that when we incorporate the R22 prior in joint analysis, the obtained value of $r_d$ shows a close agreement with the value obtained without R22 priors, indicating a reduction in the $r_d$ value. However, tension is still observable when we slowly incorporate the BAO with and without R22 priors. Crucially, our assessments highlight that the values of $H_0$ and $r_d$ based on low-redshift measurements align with early Planck estimates~\cite{1BAO}. We also compared our proposed cosmological model with observational data and the well-established $\Lambda$CDM paradigm. We found that our model's predictions aligned closely with the observed data from Cosmic Chronometers, indicating its ability to describe the Universe's expansion history. Additionally, we identified slight discrepancies between our model and the $\Lambda$CDM model at higher redshifts, suggesting the presence of unique features in our model at those epochs. The cosmographic results of our new cosmological model demonstrate its remarkable ability to describe the observed universe and its expansion history. The model exhibits excellent agreement with the standard $\Lambda$CDM paradigm at higher redshifts. However, the real excitement lies in its prediction of a super-accelerated expansion in the distant future, in contrast to the de Sitter phase predicted by $\Lambda$CDM. This finding implies the presence of dark energy driving the accelerated expansion. Our analysis of the evolution of cosmological parameters in our proposed model reveals intriguing departures from the standard $\Lambda$CDM model. The jerk parameter, characterizing the rate of change of acceleration with respect to time, demonstrates a notable increase at lower redshifts compared to the $\Lambda$CDM prediction. This deviation suggests a departure from the standard behavior and hints at potential dynamics not captured by the traditional model. The statefinder diagnostic illustrates a complex journey through various dark energy domains, shedding light on the richness of the universe's dynamics within our model. The back-and-forth movement across fixed $\Lambda$CDM points emphasizes the intricate interplay between different phases of dark energy domination. The consistent lower values of \(O_{m}\) compared to \( \Omega_{m0} \) across all redshifts suggest that the proposed models reside in the phantom region. In our ongoing exploration of various cosmological models, we employ both the Akaike Information Criterion (AIC) and the Bayesian Information Criterion (BIC) as tools for model comparison. The AIC is represented as follows \cite{71BAO} : $\mathrm{AIC} = -2 \ln(\mathcal{L}_{\text{max}}) + 2k + \frac{2k(2k+1)}{N_{\text{tot}} - k - 1}$. Here, \(\mathcal{L}_{\max}\) denotes the maximum likelihood of the data, encompassing the entire dataset without the R22 prior. The parameters include \(N_{\text{tot}}\), representing the total number of data points (in our case, \(N_{\text{tot}} = 273\)), and \(k\), indicating the number of parameters. For large \(N_{\text{tot}}\), this equation simplifies to: $\mathrm{AIC} \simeq -2 \ln(\mathcal{L}_{\max}) + 2k$, which is the standard form of the AIC criterion \cite{71BAO}. Conversely, the Bayesian Information Criterion is expressed as \cite{72BAO}: $\mathrm{BIC} = -2 \ln(\mathcal{L}_{\max}) + k \ln N_{\text{tot}}$. Using these criteria, we calculate the AIC and BIC for the standard \(\Lambda\)CDM and new purposed model. The computed values for \(\Lambda\)CDM and the New Model are respectively AIC = $[264.538, 271.326]$ and BIC = $[264.763, 272.166]$. Despite the \(\Lambda\)CDM model exhibiting the best fit with the lowest AIC, our combined AIC and BIC results offer support to all the tested models, indicating that none of them can be dismissed based solely on the existing data. In evaluating the New Cosmological Model relative to \(\Lambda\)CDM, we acknowledge that \(\Lambda\)CDM is encompassed within both proposed extensions, differing by 5 degrees of freedom. This difference permits the use of standard statistical tests. The benchmark for comparison is the reduced chi-square statistic, defined as \( \chi_{\text{red}}^{2} = \chi^{2} / \text{Dof} \), where Dof represents the degrees of freedom of the model, and \( \chi^{2} \) denotes the weighted sum of squared deviations. With an equal number of runs for the three models, the statistic approximates 1, expressed as: $\left( \frac{\chi^{2}}{Dof_{\Lambda CDM}}, \frac{\chi^{2}}{Dof_{New Model}} \right) \approx \{0.949771, 0.955615\}$. This comparative analysis provides valuable insights into the goodness of fit for each model, with values close to 1 indicating a satisfactory alignment with the observed data. These findings suggest that the New Cosmological Model could potentially mimic established models. However, further studies are necessary to thoroughly assess its validity. The model has passed some crucial tests, as detailed in the corresponding manuscript. The manuscript mainly focuses on utilizing late-time datasets to obtain essential cosmological parameters and to study cosmography, diagnostics and statistical analysis.

\section*{Data Availability}             
No new data were generated in support of this research.

\section*{Conflict of Interest} 
The authors declare no conflict of interest.

\section*{Acknowledgement}
 The author S. K. Maurya is thankful for continuous support and encouragement from the administration of University of Nizwa. 
 
\bibliographystyle{ieeetr}
\bibliography{main,mybib}
    
\end{document}